\documentstyle[preprint,aps]{revtex}

\begin{document}

\title{Energy and Momentum of a Class of Rotating Gravitational Waves}

\author{M. Sharif \thanks{Present Address: Department of Physics,
Hanyang University, Seoul 133-791, KOREA}\\ Department of
Mathematics, Punjab University, Quaid-e-Azam Campus Lahore-54590,
PAKISTAN, e\_-mail: hasharif@yahoo.com}

\maketitle

\begin{abstract}
{\it We calculate energy and momentum for a class of cylindrical
rotating gravitational waves using Einstein and Papapetrou's
prescriptions. It is shown that the results obtained are reduced
to the special case of the cylindrical gravitational waves already
available in the literature.}
\end{abstract}

\newpage

\section{Introduction}

The notion of energy has been one of the most thorny and important
problems in Einstein's theory of General Relativity (GR). There
have been many attempts [1,2] to get a well defined expression for
local or quasi-local energy and momentum. However, there is still
no generally accepted definition known. As a result, different
people have different points of view. Cooperstock [3] argued that
in GR, energy and momentum are localized in regions of the
non-vanishing energy and momentum tensor and consequently
gravitational waves are not carriers of energy and momentum in
vacuum. The gravitational waves, by definition, have zero
stress-energy tensor. Thus the existence of these waves was
questioned. However, the theory of GR indicates the existence of
gravitational waves as solutions of Einstein's field equations
[4]. Infact this problem arises because energy is not well defined
in GR.

The problem for gravitational waves was resolved by Ehlers and
Kundt [5], Pirani [6] and Weber and Wheeler [7] by considering a
sphere of test particles in the path of the waves. They showed
that these particles acquired a constant momentum from the waves.
Qadir and Sharif [8] presented an operational procedure, embodying
the same principle, to show that gravitational waves impart a
momentum. Rosen [9] investigated whether or not cylindrical
gravitational waves have energy and momentum. He used the
energy-momentum pseudo tensors of Einstein and Landau Lifshitz and
carried out calculations in cylindrical polar coordinates.
However, he arrived at the conclusion that the energy and momentum
density components vanish. These results supported the conjecture
of Scheidegger [10] that physical system cannot radiate
gravitational energy. Later, he pointed out [11] that the energy
and momentum densities turn out to be non-vanishing and reasonable
if the calculations are performed in Cartesian coordinates. Rosen
and Virbhadra [12] explicitly evaluated these quantities in the
Einstein's prescription by using Cartesian coordinates and found
them finite and well defined. Virbhadra [13] then used Tolman,
Landau-Lifshitz and Papapetrou's prescriptions to evaluate the
energy and momentum densities and found that the same results turn
out in all these prescriptions.

In this paper we use Einstein and Papapetrou's prescriptions to
evaluate energy and momentum densities to a class of cylindrical
rotating gravitational waves. As we shall see from the analysis
given in the paper, when rotation is included, the problem is
considerably complicated. We find that the results obtained by
these two prescriptions are not exactly the same. However, it is
shown that they both reduce to the same result for a special case
of cylindrical gravitational waves. In the next section, we shall
describe the class of rotating cylindrical gravitational waves. In
sections three and four, we evaluate energy and momentum using
Einstein and Papapetrou's prescriptions respectively. Finally, we
shall discuss the results.

\section{A Class of Rotating Cylindrical Gravitational Waves}

A class of solutions of the gravitational field equations
describing vacuum spacetimes outside rotating cylindrical sources
is given by the line element of the form [14]
\begin{equation}
ds^2=e^{2\gamma-2\psi}(dt^2-d\rho^2)-\mu^2e^{-2\psi}(\omega
dt+d\phi)^2-e^{2\psi}dz^2,
\end{equation}
in the cylindrical coordinates $(\rho,\phi,z)$. Here the metric
functions $\gamma,\mu,\psi$ and $\omega$ depend on the coordinates
$t$ and $\rho$ only. In general the spacetimes have two Killing
vectors, one is associated with the invariant translations along
the symmetry axis, $\xi_{(z)}=\partial z$, and the other is
associated with the invariant rotations about the axis,
$\xi_{(\phi)}=\partial\phi$. Obviously, these Killing vectors are
orthogonal. When $\omega=0$, the metric represents spacetimes
without rotation, in which the polarization of gravitational waves
has only one degree of freedom and the direction of polarization
is fixed [4]. It is to be noticed that if we take $\omega=0$ and
$\mu=\rho$, the above metric reduces to a special case of
cylindrical gravitational waves [7]. Einstein's vacuum field
equations for the metric form (1) are given by
\begin{equation}
(\mu\psi_v)_u+(\mu\psi_u)_v=0,
\end{equation}
\begin{equation}
\mu_{uv}-\frac{l^2}{8}\mu^{-3}e^{2\gamma}=0,
\end{equation}
\begin{equation}
\omega_v-\omega_u=l\mu^{-3}e^{2\gamma},
\end{equation}
\begin{equation}
\gamma_u=\frac{1}{2\mu_u}(\mu_{uu}+2\mu\psi^2_u),
\end{equation}
\begin{equation}
\gamma_v=\frac{1}{2\mu_v}(\mu_{vv}+2\mu\psi^2_v),
\end{equation}
where $\psi_u=\frac{\partial\psi}{\partial u}$, etc. The
subscripts $u=t-\rho$ and $v=t+\rho$ are retarded and advanced
times respectively. Here $l$ is a constant length characteristic
of the rotation of the system which is positive and is
specifically attributed with rotating gravitational waves. For
$l=0$ we have $\omega=\omega(t)$ from Eq.(4) and
$\mu_{tt}=\mu_{\rho\rho}$ from Eq.(3). A simple transformation to
a rotating frame reduces the waves to non-rotating generalized
Beck spacetimes which have been studied by many authors [15,16].

In order to have meaningful results in the prescription of
Einstein and Papapetrou, it is necessary to transform the metric
in Cartesian coordinates. Let us transform the metric in Cartesian
coordinates by using
\begin{equation}
x=\rho\cos\phi,\quad y=\rho\sin\phi.
\end{equation}
The corresponding metric in these coordinates will become
\begin{equation}
ds^2=e^{2\gamma-2\psi}(dt^2-\frac{1}{\rho^2}(xdx+ydy)^2)-\mu^2e^{-2\psi}(\omega
dt+\frac{1}{\rho^2}(xdy-ydx))^2-e^{2\psi}dz^2.
\end{equation}

\section{Energy and Momentum in Einstein's Prescription}

The energy-momentum complex of Einstein [17] is given by
\begin{equation}
\Theta_a^b=\frac{1}{16\pi}H^{bc}_{a\;,c},
\end{equation}
where
\begin{equation}
H_a^{bc}=\frac{g_{ad}}{\sqrt{-g}}[-g(g^{bd}g^{ce}-g^{cd}g^{be})]_{,e},\quad
a,b,c,d,e=0,1,2,3.
\end{equation}
$\Theta_0^0$ is the energy density, $\Theta_0^a$ are the momentum
density components, and $\Theta_a^0$ are the components of energy
current density. The Einstein energy-momentum satisfies the local
conservation laws
\begin{equation}
\frac{\partial\Theta_a^b}{\partial x^b}=0.
\end{equation}
The required non-vanishing components of $H_a^{bc}$ are
\begin{equation}
H_0^{01}=\frac{1}{\mu\rho^3}(\mu^2x-2\mu\mu_\rho\rho x
+\rho^2xe^{2\gamma}-\mu^2\omega^2\rho^2x+2\mu^2\omega\dot{\gamma}\rho^2y
-\mu^4\omega\omega_\rho\rho xe^{-2\gamma}),
\end{equation}
\begin{equation}
H_0^{02}=\frac{1}{\mu\rho^3}(\mu^2y-2\mu\mu_\rho\rho y
+\rho^2ye^{2\gamma}-\mu^2\omega^2\rho^2y-2\mu^2\omega\dot{\gamma}\rho^2x
-\mu^4\omega\omega_\rho\rho ye^{-2\gamma}),
\end{equation}
\begin{equation}
H_1^{01}=-\frac{1}{\rho^4}(2\mu\dot{\gamma}\rho y^2+2\dot{\mu}\rho
x^2-\mu^3\omega_\rho xye^{-2\gamma}),
\end{equation}
\begin{equation}
H_1^{02}=\frac{1}{\rho^4}(\mu\omega\rho^3+2\mu\dot{\gamma}\rho
xy+\mu^3\omega_\rho y^2e^{-2\gamma}),
\end{equation}
\begin{equation}
H_2^{01}=\frac{1}{\rho^4}(2\mu\dot{\gamma}\rho xy-2\dot{\mu}\rho
xy-\mu\omega\rho^3-\mu^3\omega_\rho x^2e^{-2\gamma}),
\end{equation}
\begin{equation}
H_2^{02}=-\frac{1}{\rho^4}(2\mu\dot{\gamma}\rho x^2+2\dot{\mu}\rho
y^2+\mu^3\omega_\rho xye^{-2\gamma}),
\end{equation}
\begin{equation}
H_3^{03}=-\frac{2}{\rho}(\mu\dot{\gamma}-2\mu\dot{\psi}+\dot{\mu}),
\end{equation}
\begin{equation}
H_0^{12}=-\mu\omega_\rho+2\mu\omega\gamma_\rho-2\mu_\rho\omega,
\end{equation}
\begin{equation}
H_0^{03},\quad H_1^{03},\quad H_2^{03},\quad H_3^{01},\quad
H_3^{02},\quad H_0^{23},\quad H_0^{31}=0.
\end{equation}
Using Eqs.(12-20) in Eq.(9), we obtain energy and momentum
densities in Einstein's prescription
\begin{eqnarray}
\Theta_0^0=\frac{1}{16\pi\mu^2\rho^3}[\mu^2(-\mu+\mu_\rho\rho-2\mu_{\rho\rho}\rho^2
-\mu\omega^2\rho^2-2\mu\omega\omega_\rho\rho^3-\mu_\rho\omega^2\rho^3)\\\nonumber
+\rho^2(\mu+2\mu\gamma_\rho\rho-\mu_\rho\rho)e^{2\gamma}-\mu^4\rho^2(\mu\omega_\rho^2
+\mu\omega\omega_{\rho\rho}\\\nonumber
-2\mu\omega\omega_\rho\gamma_\rho+3\mu_\rho\omega\omega_\rho)e^{-2\gamma}],
\end{eqnarray}
\begin{eqnarray}
\Theta_1^0=\frac{1}{16\pi\rho^5}[2\mu\rho^2\dot{\gamma}x-6\dot{\mu}x^3-2\dot{\mu}\rho^2x
-2\dot{\mu_\rho}\rho x^3 -\mu\omega^2\rho y\\\nonumber -\mu^2\rho
y(\mu\omega_\rho
-\mu\omega_{\rho\rho}\rho+2\mu\omega_\rho\gamma_\rho\rho-3\mu\mu_\rho\omega_\rho\rho)
e^{-2\gamma}],
\end{eqnarray}
\begin{eqnarray}
\Theta_2^0=\frac{1}{16\pi\rho^4}[2\mu\rho\dot{\gamma}y-2\dot{\mu_\rho}\rho^2y
+\mu\omega\rho
x-\mu\omega_{\rho}\rho^2x-\mu_\rho\omega\rho^2x\\\nonumber
+\mu^2x(\mu\omega_\rho-\mu\omega_{\rho\rho}\rho
+2\mu\omega_\rho\gamma_\rho\rho-3\mu_\rho\omega_\rho\rho)e^{-2\gamma}],
\end{eqnarray}
\begin{eqnarray}
\Theta_0^1=\frac{1}{16\pi\mu^2\rho^3}[\mu^2(-\dot{\mu}x+2\dot{\mu_\rho}\rho
x+2\mu\omega\dot{\omega}\rho^2x-2\mu\omega\ddot{\gamma}\rho^2y
+2\mu\omega_\rho\gamma_\rho\rho^2y\\\nonumber
-\mu\omega_{\rho\rho}\rho^2y+2\mu\omega\gamma_{\rho\rho}\rho^2y
+\dot{\mu}\omega^2\rho^2x+2\mu_\rho\omega\gamma_\rho\rho^2y
-3\mu_\rho\omega_\rho\rho^2y\\\nonumber
-2\mu_{\rho\rho}\omega\rho^2y)-\rho^2x(2\mu\dot{\gamma}-\dot{\mu})e^{2\gamma}
+\mu^4\rho x(\mu\dot{\omega}\omega_\rho\\\nonumber
+\mu\omega\dot{\omega_\rho}
-3\mu\omega\omega_\rho\dot{\gamma}+3\dot{\mu}\omega\omega_\rho)e^{-2\gamma}],
\end{eqnarray}
\begin{eqnarray}
\Theta_0^2=\frac{1}{16\pi\mu^2\rho^3}[\mu^2(2\dot{\mu_\rho}\rho y
+\mu\omega_{\rho\rho}\rho^2y-2\mu\omega_\rho\gamma_\rho\rho^2y
-2\mu\omega\gamma_{\rho\rho}\rho^2y+2\mu\omega\dot{\omega}\rho^2y\\\nonumber
+2\mu\dot{\omega}\dot{\gamma}\rho^2x
+2\mu\omega\ddot{\gamma}\rho^2x
+\dot{\mu}\omega^2\rho^2y+2\dot{\mu}\omega\dot{\gamma}\rho^2
x+3\mu_\rho\omega_\rho\rho^2y\\\nonumber
-2\mu_\rho\omega\gamma_\rho\rho^2y
+2\mu_{\rho\rho}\omega\rho^2y)-\rho^2y(2\mu\dot{\gamma}-\dot{\mu})e^{2\gamma}
+\mu^4\rho
y(\mu\dot{\omega}\omega_\rho\\\nonumber+\mu\omega\dot{\omega_\rho}
-2\mu\omega\omega_\rho\dot{\gamma}
+3\dot{\mu}\omega\omega_\rho)e^{-2\gamma}],
\end{eqnarray}
\begin{equation}
\Theta_0^3=\Theta_3^0=0.
\end{equation}
Now for $\omega=0$ and $\mu=\rho$, Eqs.(21)-(26) become
\begin{equation}
\Theta_0^0=\frac{1}{8\pi}e^{2\gamma}(\psi_\rho^2+\psi_t^2),
\end{equation}
\begin{equation}
\Theta_1^0=\frac{1}{4\pi\rho}x\psi_\rho\psi_t,
\end{equation}
\begin{equation}
\Theta_2^0=\frac{1}{4\pi\rho}y\psi_\rho\psi_t,
\end{equation}\begin{equation}
\Theta_0^1=-e^{2\gamma}\Theta_1^0,
\end{equation}
\begin{equation}
\Theta_0^2=-e^{2\gamma}\Theta_2^0,
\end{equation}
\begin{equation}
\Theta_0^3=\Theta_3^0=0.
\end{equation}
These are the energy and momentum densities of cylindrical
gravitational waves given by Rosen and Virbhadra [12].

\section{Energy and Momentum in Papapetrou's Prescription}

The symmetric energy-momentum complex of Papapetrou [18] is given
by
\begin{equation}
\Omega^{ab}=\frac{1}{16\pi}N^{abcd}_{\quad\;\,,cd},
\end{equation}
where
\begin{equation}
N^{abcd}=\sqrt{-g}(g^{ab}\eta^{cd}-g^{ac}\eta^{bd}+g^{cd}\eta^{ab}-g^{bd}\eta^{ac}),
\end{equation}
and $\eta^{ab}$ is the Minkowski spacetime. The energy-momentum
complex satisfies the local conservation laws
\begin{equation}
\frac{\partial\Omega^{ab}}{\partial x^b}=0.
\end{equation}
The locally conserved energy-momentum complex $\Omega^{ab}$
contains contributions from the matter, non-gravitational and
gravitational fields. $\Omega^{00}$ and $\Omega^{0a}$ are the
energy and momentum (energy current) density components. The
required non-vanishing components of $N^{abcd}$ are given as
\begin{equation}
N^{0101}=\frac{1}{\mu\rho^3}(\mu^2\rho^2+\mu^2x^2-\mu^2\rho^2\omega^2y^2
+\rho^2y^2e^{2\gamma}),
\end{equation}
\begin{equation}
N^{0102}=\frac{1}{\mu\rho^3}(\mu^2xy+\mu^2\omega^2\rho^2xy-\rho^2xye^{2\gamma}),
\end{equation}
\begin{equation}
N^{0202}=\frac{1}{\mu\rho^3}(\mu^2\rho^2+\mu^2y^2-\mu^2\rho^2\omega^2x^2
-\rho^2x^2e^{2\gamma}),
\end{equation}
\begin{equation}
N^{0303}=\frac{\mu}{\rho}(1+e^{2\gamma-4\psi}),
\end{equation}
\begin{equation}
N^{0121}=-\frac{1}{\rho}(\mu\omega x),
\end{equation}
\begin{equation}
N^{0122}=-\frac{1}{\rho}(\mu\omega y),
\end{equation}
Substituting Eqs.(36-41) in Eq.(33), we have energy and momentum
density components in Papapetrou's prescription
\begin{eqnarray}
\Omega^{00}=\frac{1}{16\pi\mu^2\rho^3}[\mu^2(-\mu+\mu_\rho\rho-2\mu_{\rho\rho}\rho^2
-\mu\omega^2\rho^2-2\mu\omega\omega_\rho\rho^3-\mu_\rho\omega^2\rho^3)\\\nonumber
+(\mu\rho^2+2\mu\gamma_\rho\rho^3-\mu_\rho\rho^3)e^{2\gamma}],
\end{eqnarray}
\begin{eqnarray}
\Omega^{01}=\frac{1}{16\pi\mu^2\rho^3}[\mu^2(-\dot{\mu}x+2\dot{\mu_\rho}\rho
x+\mu\omega y-\mu\omega_\rho\rho
y-\mu\omega_{\rho\rho}\rho^2y\\\nonumber
+2\mu\omega\dot{\omega}\rho^2x+\dot{\mu}\omega^2\rho^2x-\mu_\rho\omega\rho
y-2\mu_\rho\omega_\rho\rho^2y\\\nonumber
-\mu_{\rho\rho}\omega\rho^2y)-\rho^2x(2\mu\dot{\gamma}-\dot{\mu})e^{2\gamma}],
\end{eqnarray}
\begin{eqnarray}
\Omega^{02}=\frac{1}{16\pi\mu^2\rho^3}[\mu^2(\dot{\mu}y+2\dot{\mu_\rho}\rho
y+2\mu\omega\dot{\omega}\rho^2y+\dot{\mu}\omega^2\rho^2y-\mu\omega
x\\\nonumber +\mu\omega_\rho\rho x+\mu\omega_{\rho\rho}\rho^2x
+\mu_\rho\omega \rho x+2\mu_\rho\omega_\rho\rho^2x\\\nonumber
+\mu_{\rho\rho}\omega\rho^2x)-\rho^2
y(2\mu\dot{\gamma}-\dot{\mu})e^{2\gamma}],
\end{eqnarray}
\begin{equation}
\Omega^{03}=\Omega^{30}=0.
\end{equation}
We see that for $\omega=0$ and $\mu=\rho$, Eqs.(42)-(45) yield
\begin{equation}
\Omega^{00}=\frac{1}{8\pi}e^{2\gamma}(\psi_\rho^2+\psi_t^2),
\end{equation}
\begin{equation}
\Omega^{01}=-\frac{1}{4\pi\rho}x\psi_\rho\psi_te^{2\gamma},
\end{equation}
\begin{equation}
\Omega^{02}=-\frac{1}{4\pi\rho}y\psi_\rho\psi_te^{2\gamma},
\end{equation}
\begin{equation}
\Omega^{03}=0.
\end{equation}

These turn out to be energy and momentum density components for
cylindrical gravitational waves given by Virbhadra [13].

\section{Discussion}

We have evaluated energy and momentum density components for a
class of rotating cylindrical gravitational waves by using
prescriptions of Einstein and Papapetrou. It can be seen that the
energy and momentum densities for a class of rotating
gravitational waves are finite and well defined in both the
prescriptions. It follows from Eqs.(21-26) and (42-45) that though
the energy-momentum complexes of Einstein and Papapetrou are not
exactly the same but are similar upto certain terms. However, it
is interesting to note from Eqs.(27-32) and (46-49) that both the
results reduce to the same energy and momentum densities of a
special case of cylindrical gravitational waves as given in
[12,13].
\newpage

\begin{description}
\item  {\bf Acknowledgment}
\end{description}

The author would like to thank Prof. Chul H. Lee for his
hospitality at the Department of Physics and Korea Scientific and
Engineering Foundation (KOSEF) for postdoc fellowship at Hanyang
University Seoul, KOREA.

\vspace{2cm}

{\bf \large References}

\begin{description}

\item{[1]} Penrose, R.: Proc. Roy. Soc. London {\bf
A381}(1982)53;\\ Misner, C.W. and Sharp, D.H. Phys. Rev. {\bf
136}(1964)B571.

\item{[2]} Bondi, H.: Proc. Roy. Soc. London {\bf A427}(1990)249.

\item{[3]} Cooperstock, F.I.: Found. Phys. {\bf 22}(1992)1011; in {\it Topics
in Quantum Gravity and Beyond: Papers in Honor of L. Witten} eds.
Mansouri, F. and Scanio, J.J. (World Scientific, Singapore, 1993)
201; in {\it Relativistic Astrophysics and Cosmology}, eds.
Buitrago et al (World Scientific, Singapore, 1997)61; Annals Phys.
{\bf 282}(2000)115.

\item{[4]} Misner, C.W., Thorne, K.S. and Wheeler, J.A.: {\it Gravitation} (W.H. Freeman,
San Francisco, 1973).

\item{[5]} Ehlers, J. and Kundt, W.: {\it Gravitation: An Introduction to Current
Research}, ed. L. Witten (Wiley, New York, 1962)49.

\item{[6]} Pirani, F.A.E.: {\it Gravitation: An Introduction to Current
Research}, ed. L. Witten (Wiley, New York, 1962)199.

\item{[7]} Weber, J. and Wheeler, J.A.: Rev. Mod. Phy. {\bf 29}(1957)509;\\
Weber, J.: {\it General Relativity and Gravitational Waves},
(Interscience, New York, 1961).

\item{[8]} Qadir, A. and Sharif, M.: Physics Letters {\bf A
167}(1992)331.

\item{[9]} Rosen, N: Helv. Phys. Acta. Suppl. {\bf 4}(1956)171.

\item{[10]} Scheidegger, A.E.: Rev. Mod. Phys. {\bf 25}(1953)451.

\item{[11]} Rosen, N.: Phys. Rev. {\bf 291}(1958)110.

\item{[12]} Rosen, N. and Virbhadra, K.S.: Gen. Rel. and Grav. {\bf
26}(1993)429.

\item{[13]} Virbhadra, K.S.: Pramana-J. Phys. {\bf 45}(1995)215.

\item{[14]} Mashhoon, Bahram , McClune, James C. and Quevedo, Hernando: Class.
Quantum Grav.{\bf 17}(2000)533.

\item{[15]} Kramer, D., Stephani, H. , MacCallum, M. and Herlt, E.: {\it Exact
Solutions of Einstein's Field Equations} (Cambridge University
Press, Cambridge, 1980).

\item{[16]} Verdaguer, E.: Phys. Rep. {\bf 229}(1993)1;\\
Griffths, J.B.: {\it Colliding Plane Waves in General Relativity}
(Oxford University Press, Oxford, 1991);\\ Alekseev, G.A. and
Griffths, J.B.: Class. Quantum Grav. {\bf 13}(1996)2191.

\item{[17]} M$\ddot{o}$ller, C.: Ann. Phys. (NY) {\bf 4}(1958)347; {\bf
12}(1961)118.

\item{[18]} Papapetrou, A.: Proc. R. Irish. Acad. {\bf A52}(1948)11.

\end{description}

\end{document}